# Observer study-based evaluation of TGAN architecture used to generate oncological PET images

Roberto Fedrigo[*a,b], Fereshteh Yousefirizi[a], Ziping Liu[c,d], Abhinav K. Jha[c,d],
Robert V. Bergen[e], Jean-Francois Rajotte[e], Raymond T. Ng[e], Ingrid Bloise[f],
Sara Harsini[f], Dan J. Kadrmas[g], Carlos Uribe[a,f,h], Arman Rahmim[a,b,f,h]

[a]Dept. of Integrative Oncology, BC Cancer Research Institute, Vancouver, BC Canada
[b]Dept. of Physics and Astronomy, Univ. of British Columbia, Vancouver BC Canada
[c]Dept. of Biomedical Engineering, Washington Univ. in St. Louis, St. Louis, MO USA
[d]Mallinckrodt Institute of Radiology, Washington Univ. School of Med., St. Louis, MO USA
[e]Univ. of British Columbia Data Science Institute, 6339 Stores Road, Vancouver, BC Canada
[f]Dept. of Functional Imaging, BC Cancer, Vancouver, BC Canada
[g]Dept. Of Radiology and Imaging Sciences, University of Utah, Salt Lake City, UT USA
[h]Dept. of Radiology, Univ. of British Columbia, Vancouver, BC Canada

**ABSTRACT**

The application of computer-vision algorithms in medical imaging has increased rapidly in recent years. However, algorithm training is challenging due to limited sample sizes, lack of labeled samples, as well as privacy concerns regarding data sharing. To address these issues, we previously developed (Bergen et al. 2022) a synthetic PET dataset for Head & Neck (H&N) cancer using the temporal generative adversarial network (TGAN) architecture and evaluated its performance segmenting lesions and identifying radiomics features in synthesized images. In this work, a two-alternative forced-choice (2AFC) observer study was performed to quantitatively evaluate the ability of human observers to distinguish between real and synthesized oncological PET images. In the study eight trained readers, including two board-certified nuclear medicine physicians, read 170 real/synthetic image pairs presented as 2D-transaxial using a dedicated web app. For each image pair, the observer was asked to identify the "real" image and input their confidence level with a 5-point Likert scale. P-values were computed using the binomial test and Wilcoxon signed-rank test. A heat map was used to compare the response accuracy distribution for the signed-rank test. Response accuracy for all observers ranged from 36.2% [27.9-44.4] to 63.1% [54.8-71.3]. Six out of eight observers did not identify the real image with statistical significance, indicating that the synthetic dataset was reasonably representative of oncological PET images. Overall, this study adds validity to the realism of our simulated H&N cancer dataset, which may be implemented in the future to train AI algorithms while favoring patient confidentiality and privacy protection.

## 1. INTRODUCTION

Positron emission tomography (PET) imaging is used to assess disease pathology in a wide variety of fields including neurology, cardiology, and oncology [1]. The application of computer-vision algorithms in PET has rapidly increased in recent years [2], resulting in dedicated artificial intelligence (AI) based approaches for tumor detection, segmentation, and image quantification tasks [3]. However, training state-of-the-art AI models is challenging due to limited annotated data (labels). Data sharing between institutions is one way to satisfy data requirements, but can be difficult due to privacy concerns [4]. On the other hand, public medical dataset availability is limited, particularly for PET, and those that are available may vary in quality. Additionally, some disease types may be rare, giving rise to imbalanced data.

Possible solutions to these problems include federated learning and use of synthetic images. Due to the limitations of the former including the heterogeneity of the data, ethical issues regarding the patient information leakage and lack of proper frameworks, the latter solution may be preferred. For this reason, a robust generative method to create synthetic data is highly sought after. Unfortunately, most three-dimensional (3D) image generators are extremely memory intensive and/or require additional image inputs. To address these issues, we already proposed adapting video generation techniques for 3D image generation using the temporal generative adversarial network (TGAN) architecture [5]. Due to the important role of PET imaging in the diagnosis and management of Head & Neck (H&N) cancer [6], [7], implementation of an automatic, accurate, and robust gross tumor volume segmentation and radiomics analysis is in high demand for effective H&N cancer management. We used the TGAN architecture to generate a synthetic data set of H&N cancer patients. This architecture provides the ability to model patient anatomy, as well as to select tumor geometry and location in the synthetic images.

To highlight the utility of our TGAN architecture for computer-vision applications, we previously trained a segmentation model using synthetic images conditioned on real tumor masks. We showed that the segmentation algorithm had similar performance for both the real and synthetic datasets (Dice scores of 0.7 and 0.65, respectively). Furthermore, radiomics features were highly correlated between the real and synthetic images, which is indicative of the realistic imaging features generated by the TGAN architecture. However, we have not yet validated the *clinical realism* of these images for more complex tasks, such as the training of algorithms which seek to perform an objective assessment of image quality (OAIQ) in PET images [8], [9]. An important goal of our synthetic dataset is that it effectively models the healthy tissue and tumor uptake observed in PET images of H&N cancer patients.

One possible approach to evaluate the realism of simulated images is using a forced-choice detection paradigm with human observers [10]. Within the forced-choice detection paradigm, the observer is presented with paired images (e.g., one real and one synthetic) and is asked to differentiate between them. It is well known that computing the probability of an observer making a correct assignment in the two-alternative forced-choice (2AFC) study is the same as computing the AUC of that observer [11], [12]. By following the mathematical treatment in Barrett et al. [9] but focusing in the context of evaluating the clinical realism of synthetic images, Liu et al. [10] showed that an ideal-observer-study-based approach provides a mechanism to quantitatively assess the similarity in distributions of the real and synthetic images. This ideal observer, while optimal, is typically very challenging to obtain in clinical studies. Another approach is to conduct a 2AFC study with expert human observers, such as physicians with multiple years of experience in reading PET scans. Due to their expertise, the physicians are best placed among all the available human observers to evaluate the clinical realism of synthetic images. Additionally, the 2AFC study provides a mechanism to quantify the performance of physicians on this task. In this study, we implement a 2AFC observer-based framework to quantitatively evaluate the clinical realism of our synthetic dataset. In future work, we will utilize confidence ratings and qualitative feedback from expert physicians to refine our TGAN architecture to or training process by adding the feedbacks as new constrains or explainability to generate more realistic PET images of H&N cancer patients.

## 2. METHODS

### 2.1 TGAN architecture

The Generative Adversarial Network used in this study is a modified version of the temporal GAN (a.k.a. TGAN), which is a deep learning approach that was originally developed to generate videos [13]. As shown in **Fig. 1**, the TGAN consists of two parts: a temporal generator ($G_0$) and an image generator ($G_1$) [5]. The temporal component generates a set of latent variables, one for each frame of the video. The image component then uses these latent variables to generate a video by transforming them into corresponding video frames. The temporal generator takes a random input ($Z_0$) and produces a temporal vector ($Z_1(t)$), which is then used by the image generator to generate frames of a video at time $t$. The video is represented as a series of frames generated by $G_1$ using the inputs $Z_0$ and $Z_1(t)$. In this context, the time steps reflect the sequence of axial slices in the Z direction. To stabilize the training process, the spectral norm of the weight parameters in each layer was constrained to be less than 1, a technique that is referred to as singular value clipping [14]. Overall, this TGAN architecture was used to generate synthetic PET images, as described in the following section.

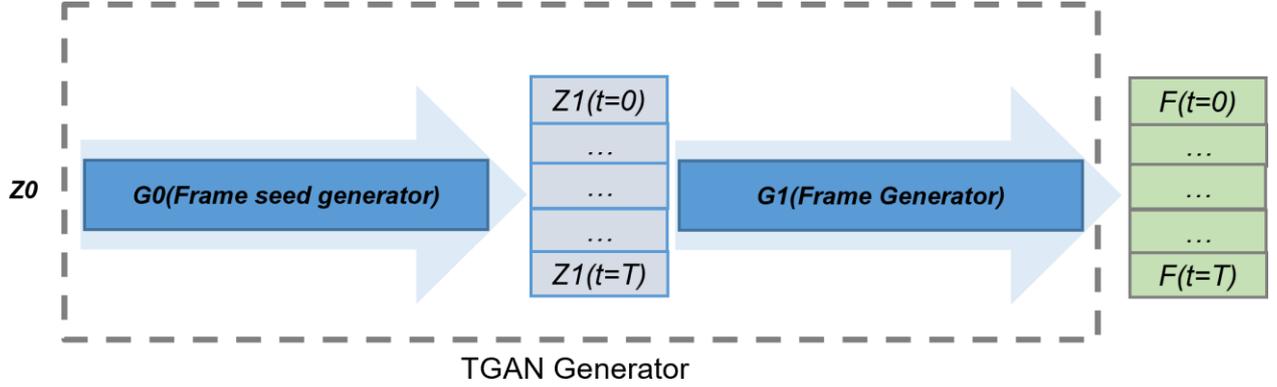

**Fig. 1:** The generator in TGAN for video creation is separated into two parts. $G_0$ generates seeds for frames by learning the temporal patterns in the video, while $G_1$ produces the individual frames.

**2.2 Generating synthetic PET images**

We utilized a publicly available dataset in The Cancer Imaging Archive (TCIA), further refined within the MICCAI 2020 Head & Neck Tumor (HECKTOR) challenge [15]; it comprises 201 cases from four centers. Each case consists of a PET image and GTVt (primary Gross Tumor Volume) mask, as well as a bounding box location (**Fig. 2a**). We used the bounding box information to crop the PET and GTVt masks to 64×64×32 volumes for input into the TGAN and segmentation networks. The in-plane (transaxial) resolution of the PET images ranged from 3.5 mm to 3.9 mm while the axial resolution was 3.7mm. After cropping, this corresponds to a minimum field of view of (224 mm × 224 mm × 118.4 mm). For unconditional TGAN, all 201 cases were used for training. For conditional TGAN, 11 cases were randomly withheld for testing. For training the segmentation neural networks, 25% of the cases were randomly withheld for testing.

**2.3 Evaluating the realism of synthetic PET images**

A 2AFC observer study was implemented using 170 real and 170 synthetic patient images (**Fig. 2b**). From each 3D image, a single tumor-present slice was randomly selected and visualized using an inverse Grayscale colormap. Pairs of real/synthetic images were imported to a dedicated 2AFC web app [10] and split into training and testing modules (n=40 and n=130, respectively). During the training session, observers were provided with the correct answer after each image set. The rationale for the training module was two-fold: a) to train/orient observers on how to use the web app interface, and b) to provide observers with feedback such that they can learn to detect any underlying differences between the real and synthetic images. During testing, the same steps as in training module were performed, except for the fact that correct answers were not released to the observers during the session.

For each image set, the web app requires the observers to select the image that they believe is real and input their confidence level on a 1-to-5 Likert scale. Optionally, the observer may scale the window level and provide qualitative feedback regarding the realism of each image set. The 2AFC study was performed with eight readers including two board-certified nuclear medicine physicians. The binomial test was applied and a 2-tailed p-value was computed, to evaluate the probability that the responses were consistent with random guess decision-making. The response accuracy distribution (i.e., using each confidence rating) was compared between observers using the Wilcoxon signed-rank test. Lastly, qualitative feedback from observers was recorded to provide qualitative feedback regarding image realism.

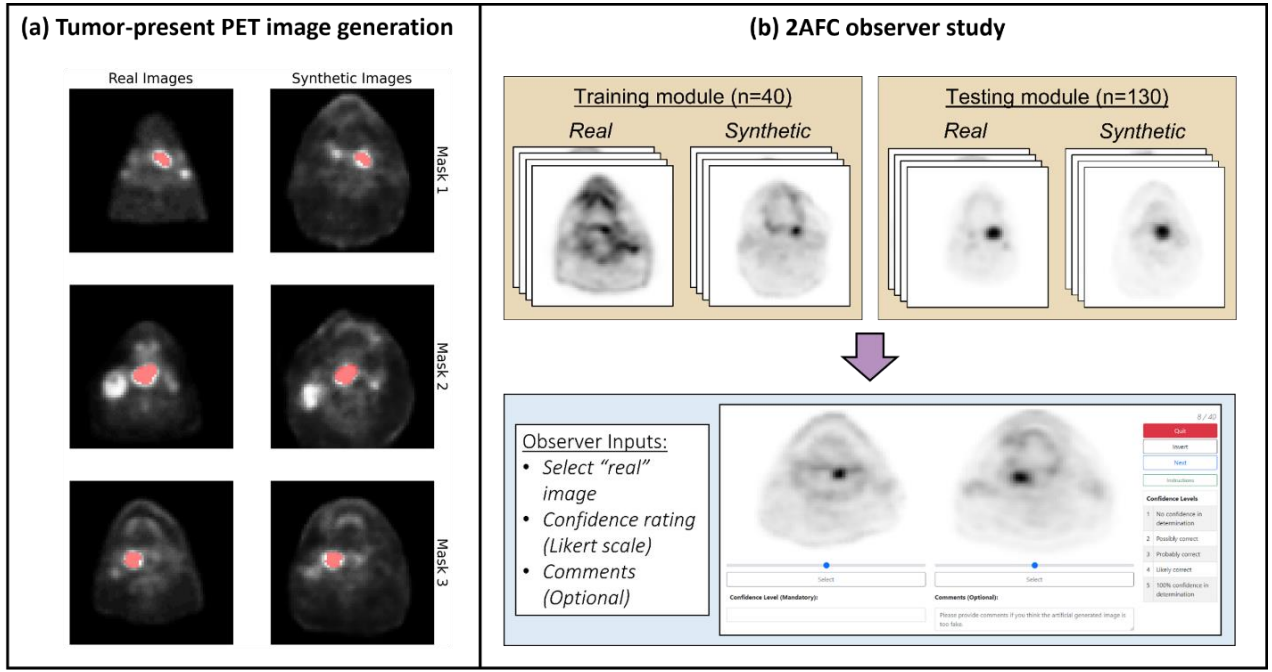

**Fig. 2: (a)** Tumor-present PET images generated to simulate Head and Neck (H&N) cancer patients. **(b)** Two-alternative forced-choice (2AFC) study split into training and testing modules (*top*) and the web app used to implement the study (*bottom*).

## 3. RESULTS

The mean response accuracy for all observers in the 2AFC study was 50.6%. Statistically significant results were obtained for reader 1 (P1) and reader 2 (P2), with p-values of 0.011 and 0.002, respectively. Notably, the statistically significant result for P2 was achieved with a less than 50% response accuracy, indicating that predictive values from their responses may be obtained by taking the inverse of their selected images. Nuclear medicine physician 1 (NMP1) had a response accuracy and confidence interval of 63.1% [54.8-71.3], which was statistically significant (p=0.003). Meanwhile, NMP2 had a response accuracy and confidence interval of 46.2% [37.6-54.7].

**Table 1:** Response accuracy, confidence interval, and p-value, for physicists (P) and nuclear medicine physicians (NMP) in the 2AFC study.

| Observer | Response Accuracy (%) | P-value |
|---|---|---|
| P1 | 61.5 [53.2 – 69.9] | 0.011* |
| P2 | 36.2 [27.9 – 44.4] | 0.002* |
| P3 | 46.2 [37.6 – 54.7] | 0.430 |
| P4 | 50.8 [42.1 – 59.4] | 0.930 |
| P5 | 57.7 [49.2 – 66.2] | 0.095 |
| P6 | 43.0 [34.6 – 51.6] | 0.136 |
| NMP1 | 63.1 [54.8 – 71.3] | 0.003* |
| NMP2 | 46.2 [37.6 – 54.7] | 0.430 |

The response accuracy for each observer and confidence rating is visualized as a heat map in **Fig. 3a**. NMP1, the highest scoring observer, had response accuracy values which monotonically increased with their confidence ratings. This indicates that the observer may have been successful in identifying features to discriminate the real and synthetic images. For P5, it can also be observed that their response accuracy increased with confidence rating, although the relationship was not monotonic. Other observers did not have a distinct relationship between their response accuracy and confidence ratings. It should be noted that 4 out of 8 observers did not select a confidence rating for five of their responses, indicating that they did not perceive any images to be clearly artificial in nature.

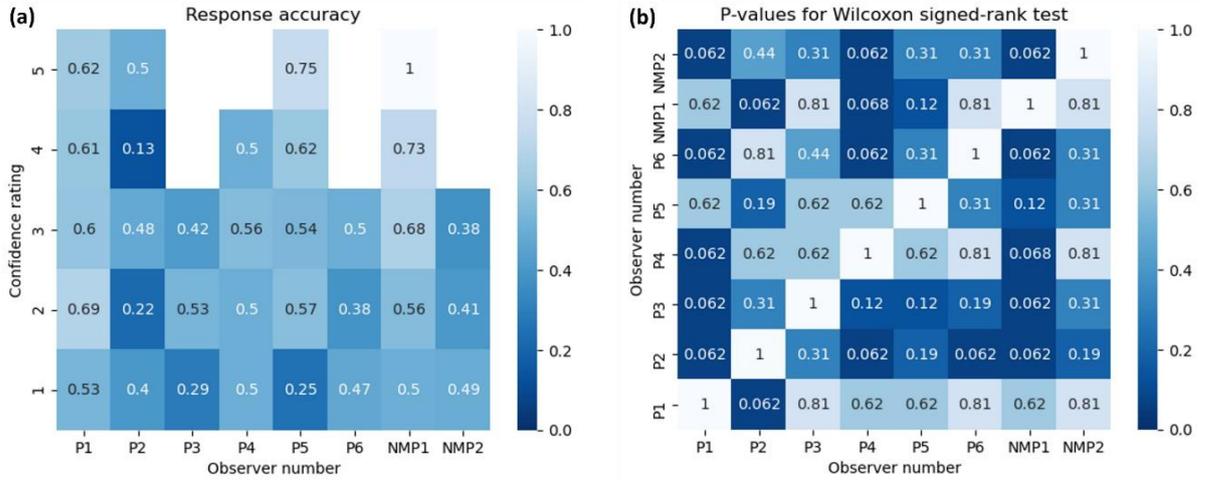

**Fig. 3: (a)** Heat map showing response accuracy for each observer and confidence rating. **(b)** Heat map of p-values for Wilcoxon signed-rank test between each combination of observers.

To compare the response accuracy distribution between observers, the Wilcoxon signed-rank test was used. This test was selected due to the inability to assume a normal distribution in our dataset. Due to the low number of categories (i.e., confidence intervals), it was not possible to achieve a p-value with 5% significance level. However, we have included a heat map of the p-values to highlight any trend that may exist between observers. Compared with other observers, P1, P4, and NMP1 had multiple p-values on the verge of significance (i.e., p=0.05-0.1). In general, P1 and NMP1 had high response accuracy compared with other observers, while P4 had near-random guessing regardless of their confidence interval (i.e., accuracy = 50-60%).

## 4. DISCUSSION

In medical imaging, there is a significant need to validate AI models for tumor detection, segmentation, and quantification, prior to introduction into a clinical setting. However, the availability of public medical datasets is quite limited and variable in terms of quality and data annotations. Conversely, most three-dimensional (3D) AI networks are extremely memory intensive or require additional image inputs. In this work, we refined and evaluated a TGAN architecture that was used to generate a synthetic dataset of H&N cancer patient images. Our TGAN-generated datasets may be freely shared while favoring patient confidentiality. This may lead to the efficient development of multi-centre studies, and lead to more direct comparison of AI algorithms in medical imaging.

In this study, we implemented a two-alternative forced-choice (2AFC) observer framework to evaluate the realism of our synthetic dataset. This provides a clear strength compared to conventional approaches, which implement a quantitative analysis and simple visual inspection to validate realism. Within this observer study, we utilized multiple observers with varying expertise (i.e., physics student to trained nuclear medicine physicians), to ensure that we have sufficient information. We performed both a quantitative and qualitative analysis of our study, to ensure that we have a conceptual understanding of our results. NMP1 provided valuable qualitative feedback which helps to understand the real and synthetic datasets. Paraphrased comments are shown below:

1. *In this image, it is challenging to delineate the contour, and its location relative to the neck is not clear. This slice appears to be low quality.*
2. *The anatomy in this image appears to be rotated. This makes it difficult to perform an assessment of its realism.*
3. *This slice does not look comparable to cases that I encounter in the clinic. If this is a real patient, it has poor resolution and noise characteristics.*

Many of the challenges experienced by NMP1 appears to be related to a lack of *contextual* information from single 2D slices. This contrasts from clinical tasks which involves reading images from multiple slices and orientations (i.e., coronal, sagittal, and transaxial). In future work, we will expand this observer study paradigm to allow observers to scroll through the entire 3D volume. To further increase the generalizability of these results, it is important to create fusions of functional and anatomical (i.e., PET/CT) information. As CT images are conventionally obtained with higher matrix sizes and resolution, this will require a substantial increase in computational time and memory. However, we expect that these modifications will result in an improvement in NMP observer accuracy, as this will more closely resemble cases encountered in their daily clinical workflow.

Overall, our study represents progress in the development and validation of realistic synthetic medical image generation. This enables creation of datasets to augment training of AI and deep learning models, while favoring patient confidentiality and privacy protection [16]. Meanwhile, we also envision that synthetic datasets may be ultimately used to further train physicians, technologists, and physicists in nuclear medicine, as also attempted in other fields [17]. Within this context, further observer studies will be needed to validate the clinical realism of these synthetic datasets before they are introduced into a clinical setting.

## 5. CONCLUSION

In this study, we performed an observer-study based evaluation of a synthetic PET dataset of H&N cancer patients. The null hypothesis could not be rejected in 5 out of 8 observers, indicating that the images appear to be relatively realistic. However, many of the physicists in the study did not have *a priori* knowledge regarding H&N cancer anatomy. This may have limited their performance during this task. We aimed to overcome this with implementation of a training module. Additionally, physician feedback indicated that images lacked context for differentiating between real vs. synthetic images. This prompts the need to implement a full 3D study to ensure geographic realism of nearby anatomy. In essence, this study helps to validate the realism of our synthetic H&N cancer dataset, which may be implemented in the future to train AI algorithms while favoring patient confidentiality and privacy protection.

## DISCLOSURES

Initial developments of this work were submitted and presented at the 2022 SPIE medical imaging conference. The previous submission highlights the implementation of the TGAN architecture for medical imaging, while this work focuses on the aspects of image quality and validation.